# Smart Surveillance: Identifying IoT Device Behaviours using ML-Powered Traffic Analysis

Reza Ryan, Napoleon Paciente, Cahil Youngs, Nickson Karie, Qian Li, Nasim Ferdosian
School of Electrical Engineering, Computing and Mathematical Sciences, Curtin University, Bentley, Australia
{reza.ryan, nickson.karie, qli, nasim.ferdosian}@curtin.edu.au
{napoleon.paciente, cahil.youngs}@student.curtin.edu.au

*Abstract*—The proliferation of Internet of Things (IoT) devices has grown exponentially in recent years, introducing significant security challenges. Accurate identification of the types of IoT devices and their associated actions through network traffic analysis is essential to mitigate potential threats. By monitoring and analysing packet flows between IoT devices and connected networks, anomalous or malicious behaviours can be detected. Existing research focuses primarily on device identification within local networks using methods such as protocol fingerprinting and wireless frequency scanning. However, these approaches are limited in their ability to monitor or classify IoT devices externally. To address this gap, we investigate the use of machine learning (ML) techniques, specifically Random Forest (RF), Multilayer Perceptron (MLP), and K-Nearest Neighbours (KNN), in conjunction with targeted network traffic monitoring to classify IoT device types and their actions. We constructed a testbed comprising an NPAT-enabled router and a diverse set of IoT devices, including smart cameras, controller hubs, home appliances, power controllers, and streaming devices. Experimental results demonstrate that IoT device and action recognition is feasible using our proposed ML-driven approach, with the RF classifier achieving the highest accuracy of 91%, while the MLP recorded the lowest accuracy at 56%. Notably, all device categories were successfully classified except for certain actions associated with security cameras, underscoring both the potential and the limitations of the proposed method.

## I. INTRODUCTION

The rising integration of Internet of Things (IoT) devices across critical infrastructure and consumer environments has substantially reshaped modern digital ecosystems [1]. This rapid growth is driven by advances in connectivity, declining device costs, improved infrastructure, and widespread consumer demand. IoT technologies are now integral across various sectors, including healthcare, industrial automation, mining, and smart homes, where they provide real-time data acquisition, operational efficiency, and automation. Despite these benefits, the massive deployment of heterogeneous IoT devices has significantly broadened the attack surface of modern networks.

IoT ecosystems are frequently targeted by cyber adversaries due to their inherent security limitations. Many devices employ minimal authentication mechanisms, lack firmware updates, or rely on default credentials, making them vulnerable to exploitation. As evidenced by high-profile incidents [2], compromised IoT devices have been used to launch large-scale distributed denial of service (DDoS) attacks and facilitate unauthorised access to critical systems. These vulnerabilities often stem from the prioritisation of cost and usability over robust security protocols during device design.

Given these challenges, accurate identification and monitoring of IoT devices are vital to mitigating risks [3]. Device-level identification enables administrators to detect anomalous behaviour, enforce access policies, and isolate compromised endpoints. While prior research has investigated techniques such as protocol fingerprinting, radio frequency scanning, and deep packet inspection for IoT device identification, these approaches typically assume access to internal network traffic or direct administrative control over network infrastructure [4]. This dependency on privileged access limits their applicability in external threat analysis and cross-network monitoring scenarios.

This study addresses the gap by proposing a machine learning (ML)-driven approach for identifying IoT devices and their executed actions through external network traffic analysis. By leveraging packet capture techniques and intrusion detection system (IDS) signatures, our approach can infer device types, manufacturers, and operational states without direct access to internal network controls. We evaluate multiple ML algorithms, including Random Forest (RF), Multilayer Perceptron (MLP), and K-Nearest Neighbours (KNN) on traffic traces collected from a controlled testbed featuring diverse IoT devices such as security cameras, smart hubs, power controllers, and streaming appliances.

The primary contributions of this paper are as follows:
- A novel external network monitoring framework for IoT device and action identification without requiring privileged network access.
- A comparative evaluation of ML algorithms (RF, MLP, KNN) for classifying IoT device types and actions using network flow features.
- A real-world IoT testbed incorporating heterogeneous devices to validate the effectiveness and practicality of the proposed approach.
- An analysis of device-specific action recognition accuracy, highlighting limitations such as low performance in certain device categories (e.g., security cameras) and directions for future improvement.

The remainder of this paper is organised as follows. Section II reviews related work on IoT device identification and traffic analysis. Section III presents our proposed methodology, including network setup and feature extraction. The ML models are described in Section IV. Section V details the device functionality identification via SNORT rules. Section VI presents experimental evaluation and results, and discusses the findings. Finally, Section VII concludes the paper and outlines potential avenues for future research.

## II. BACKGROUND AND RELATED WORK

The relationship between IoT devices and network security has been extensively explored in many studies. For this paper, it is crucial to establish a clear definition of IoT devices. Dan-Radu Berte [5] offers a functional definition of IoT devices, encompassing all devices connected to a smart home network that possess IP addresses and are capable of direct or cloud-based communication. This definition includes personal computers, laptops, and phones, which are typically not considered IoT devices. However, it is important to recognise that these devices contribute to an IoT network, not as IoT devices themselves, but as controllers [7].

IoT devices can be connected through many different means, such as Wi-Fi, Bluetooth, Zigbee and LoRaWAN. Some of these connections, such as LoRaWAN and Zigbee, are purposely designed for IoT and use their own infrastructure and bandwidth [8]. These connection types are out of scope for this paper, as we are specifically targeting those that can connect through pre-existing network infrastructure using Wi-Fi or Bluetooth. This is to investigate the effect on network security that IoT devices can introduce.

Bajpai [9] investigated IoT device identification within LANs using fingerprinting techniques such as NetBIOS, ARP, and UPnP scanning. These methods exploit device naming conventions, MAC address patterns, and service advertisements to generate unique device signatures. While effective in controlled environments, they require direct network access and lack applicability to external or WAN-based monitoring. Similarly, De Resende [9] applied ML algorithms, including MLP, KNN, and RF to classify IoT devices using LAN traffic patterns. These studies highlight the promise of ML in network traffic analysis but remain constrained to internal monitoring scenarios.

Further work by Yongxin [10] expanded the application of ML to detect compromised IoT devices, while Sivanathan [11] developed a multi-stage inference engine combining Naïve Bayes for feature extraction with RF classification, achieving over 99% accuracy in real-time IoT device identification. Although effective, these studies focus exclusively on LAN contexts and do not examine adversarial traffic perspectives or device behaviour in compromised states.

Intrusion detection systems (IDS) have also been explored in relation to IoT device monitoring. Santos [12] reviewed signature-, anomaly-, and specification-based IDS methods, emphasising their role in detecting abnormal network behaviour. While IDS provide valuable insights into IoT-related threats, it typically functions defensively within LAN environments and are not designed to fingerprint devices from external vantage points.

Research examining IoT identification from an external or adversarial perspective remains limited. Markowsky [7], for example, demonstrated passive reconnaissance using Shodan to locate vulnerable Cayman DSL routers, supplemented by active tools such as NMAP and MASSCAN. Although informative, this work focused on specific vulnerabilities rather than generalised IoT traffic fingerprinting or behavioural analysis beyond the LAN perimeter.

A comparison of these studies is presented in Table I, summarising their approaches, scope, applied techniques, and limitations. As shown, most existing work emphasises LAN-based detection and defensive monitoring, with little exploration of external network perspectives or attacker-oriented reconnaissance methods.

TABLE I: Summary of Related Work on IoT Device Detection

| Paper | Approach | Scope | Techniques | Limitations |
|---|---|---|---|---|
| Berte [5] | Functional definition of IoT devices in smart homes | LAN | N/A | Broad definition; no traffic analysis |
| Bajpai [6] | Fingerprinting (NetBIOS, ARP, UPnP) | LAN | Protocol scanning | LAN only; no external view |
| De Resende [9] | ML classification of IoT traffic | LAN | MLP, KNN, RF | LAN-only; no attacker view |
| Yongxin [10] | ML for IoT ID & compromised detection | LAN | ML models | LAN scope; small dataset |
| Sivanathan [11] | Multi-stage ML inference (NB + RF) | LAN | Naïve Bayes, RF | LAN-only; ignores compromised states |
| Santos [12] | IDS review (signature, anomaly, specification) | LAN | IDS methods | Defensive only |
| Markowsky [7] | Passive (Shodan) & active (NMAP, MASSCAN) | WAN | Recon tools | Single vuln focus |
| **This Work** | ML-based IoT device & action identification using external traffic | WAN | RF, MLP, KNN + IDS signatures | Limited dataset size; proof-of-concept stage |

Building upon these insights and as summarised in Table I, this study addresses the identified gap by proposing a machine learning-driven framework for IoT device and action identification using external network traffic analysis. The following section details our methodology, including the experimental setup, feature extraction process, and selected ML models.

## III. METHODOLOGY

This section outlines the methodology used to identify and classify IoT devices and their actions using external network traffic and machine learning. The experimental design consists of establishing a controlled testbed, capturing network traffic, processing and labelling data, and training machine learning models on extracted features.

### A. Experimental Design

To examine the behaviours and security implications of IoT devices, the experiment aims to identify these devices and their activities through network traffic analysis. The design includes a purpose-built testbed, network traffic collection, and the application of machine learning models, supplemented by defined roles for IDS. This methodology is intended to enhance understanding of IoT device behaviour, uncover security vulnerabilities, and support the development of more effective IoT network management strategies.

*B. Testbed Setup*

The experiment was conducted within a controlled network environment featuring a router configured with IPv4 Network and Port Address Translation (NPAT), following RFC 2663 specifications. NPAT was selected due to its prevalence in typical home and enterprise networks. IPv4 was used instead of IPv6, consistent with global adoption trends [13]. This setup enables realistic traffic capture for both training machine learning models and leveraging IDS for device action detection.

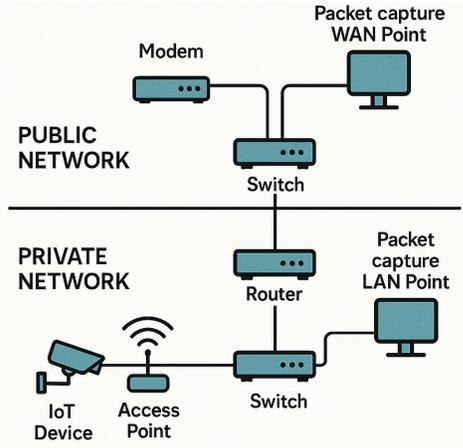

Fig. 1: Testbed Topology

Figure 1 shows the topological view of the testbed network. The network consists of two segments: the private network (LAN) and the public network (WAN). Each segment is monitored using a dedicated packet capture device. This allows for analysis of both internal device behaviour and traffic exiting the network. The NPAT router translates traffic between the LAN and WAN without any special routing rules or security restrictions, ensuring normal device operation. Introducing additional routing rules—such as port restrictions—was avoided, as these can render consumer IoT devices partially or fully non-functional.

LAN and WAN switches were incorporated to facilitate traffic capture. The WAN switch was configured to simulate scenarios like wiretapping or man-in-the-middle attacks. Windows laptops running Wireshark were used for packet capture. These devices were configured to minimise background noise, although noise levels were adjusted to simulate realistic environments when necessary. Access points supporting both 2.4 GHz and 5 GHz bands were used to connect IoT devices. Most devices operated on the 2.4 GHz band regardless of distance to the AP. Band selection was not found to impact traffic capture.

*C. Traffic Capture Methodology*

After establishing the testbed, network traffic was collected from connected IoT devices. Captured traffic was classified into two categories: passive and active.

- **Passive traffic:** Network communication initiated by a device while idle but powered and connected, such as heartbeat signals and periodic updates.
- **Active traffic:** Traffic generated during device use, including command execution, firmware updates, or user interactions.

Separating traffic into these categories ensures the machine learning models can recognise devices in both idle and operational states. For each capture, the network was initially cleared of connected devices to ensure all subsequent activity could be attributed to the device under study. Passive captures ranged from one hour to several days to capture periodic patterns. Active captures require constant user interaction through mobile applications or physical device interaction.

*D. Dataset and Feature Extraction*

The dataset used in this study was generated entirely from the testbed, as no public datasets met the study's requirements. Captured raw traffic was stored in PCAP format and transformed into structured traffic flows. These flows represent communication sessions between endpoints and form the basis for feature extraction. Once the data is converted into traffic flows, it is further segmented into distinct features. Using CICFlowMeter and custom Python scripts, we extracted 63 flow- and packet-level features, spanning traffic timing, size metrics, TCP flag counts, and window characteristics. A summary of extracted features is provided in Table II.

TABLE II: Summary of Extracted Features

| Feature Category | Example Features |
| --- | --- |
| Flow-level Metrics | Flow duration, Flow bytes/s, Flow packets/s, IAT mean/std/max/min |
| Packet-level Metrics | Packet length (min/max/mean/std/variance), Header length (fwd/bwd) |
| TCP Flags | SYN, ACK, FIN, RST, PSH flag counts |
| Bulk/Window Features | Bwd bytes/bulk avg, Subflow bytes (fwd/bwd), Init window bytes |
| Timing and Activity | Active/idle times (mean/std/max/min), Down/Up ratio |

Because the machine learning models are supervised, the dataset requires labelling. Instead of labelling by specific device models, which often exhibited similar traffic patterns, devices were grouped by functional categories: Each IoT device is categorised and labelled as either Surveillance, Hub, Energy Management, Appliance, Streaming Devices, or Non-IoT Devices. Surveillance includes devices that stream live video across the network, such as security cameras and smart doorbells. Hubs refer to IoT controller hubs and smart home centers, which are often multi-functional devices like Amazon Echo or Google Home. Energy Management encompasses devices that regulate power consumption, such as smart lights and smart power plugs. Appliance refers to IoT devices intended for household management or usage, including smart vacuums and air conditioners. Streaming Devices are entertainment-oriented IoT devices used for accessing streaming services, such as Google Chromecast and Fetch TV boxes. Non-IoT Devices cover any non-IoT devices within the network, including network noise from networking equipment, as well as computers and phones. These categories ensured better generalisation and reduced redundancy in the model training process.

## IV. MACHINE LEARNING MODELS

As discussed in the previous sections, machine learning techniques have become increasingly vital in cybersecurity for identifying and classifying IoT devices based on their network traffic patterns [14]. Due to the diverse and complex nature of IoT communications, traditional rule-based methods often fall short in capturing subtle behavioural signatures. Machine learning algorithms offer the ability to automatically learn discriminative features from traffic data, enabling more accurate and scalable detection of device types and activities. In this study, we evaluate three widely-used supervised learning algorithms to determine their effectiveness in classifying IoT devices and inferring their operational states within a controlled network environment. This section details the rationale, configuration, and hyperparameter selection for each algorithm used in this study.

### A. Random Forest

The RF classifier is an ensemble learning method primarily used for classification tasks. It operates by constructing multiple decision trees during training and outputs the class that is the mode of the predictions from individual trees. Due to its robustness to overfitting, ability to handle high-dimensional data, and effectiveness with relatively small datasets, RF has been successfully used in similar IoT traffic classification studies [6].

RF was selected over deep learning models due to the moderate size of the dataset. While deep learning generally offers strong performance with large-scale data, it often requires intensive tuning and longer training times, making it less practical for smaller, structured datasets such as the one generated in our testbed. The RF hyperparameters were tuned empirically, and the final configuration is shown in Table III. The model used 200 decision trees to ensure sufficient model stability and generalisation. The 'max_features' parameter was set to 63, which corresponds to the total number of features extracted, allowing the algorithm to consider the full feature set at each split. Equal feature weighting was maintained after testing several weighting schemes, as no significant accuracy gains were observed from feature prioritisation.

TABLE III: Hyperparameters for Random Forest Algorithm

| Hyperparameter | Value |
|---|---|
| Number of Decision Trees | 200 |
| Max Features per Split | 63 |
| Feature Weights | Equal |

It is noteworthy that assigning equal feature weights provided consistent accuracy across different device categories. For instance, surveillance devices typically generate frequent, small-sized packets due to constant video streaming, while smart hubs generate more sporadic yet information-rich packets. Attempts to weight features such as port numbers lower—due to the randomness introduced by NPAT—were also evaluated. However, certain ports were found to correlate with specific protocols (e.g., RTSP, HTTPS), providing critical signals for device type inference, so they were retained without adjustment.

### B. Multilayer Perceptron

The MLP classifier is a type of feedforward artificial neural network that maps input features to output classes through one or more hidden layers. In this experiment, the MLP model was designed with a single hidden layer containing 100 neurons. The ReLU (Rectified Linear Unit) activation function was chosen for its computational efficiency and ability to mitigate the vanishing gradient problem.

The model was trained using the Adaptive Moment Estimation (Adam) optimiser, which combines the advantages of RMSProp and AdaGrad to handle sparse data and adapt learning rates during training. Table IV outlines the key hyperparameters used for the MLP model. Given the limited dataset size, deeper MLP architectures were avoided to reduce the risk of overfitting. However, the selected configuration was sufficient to capture non-linear relationships in the traffic features and generalise across device categories.

TABLE IV: Hyperparameters for Multilayer Perceptron

| Hyperparameter | Value |
|---|---|
| Hidden Layers | 1 |
| Neurons per Hidden Layer | 100 |
| Activation Function | ReLU |
| Solver | Adam |

### C. K-Nearest Neighbours

KNN is a non-parametric, instance-based learning algorithm that classifies samples based on the majority class among the $k$ nearest neighbours in the feature space. Its simplicity and effectiveness in small to medium datasets make it suitable for baseline comparison. The value of $k = 5$ was selected based on cross-validation. Uniform weighting was applied, giving equal influence to all neighbours. The 'auto' setting allowed the underlying library to choose the most efficient algorithm (e.g., KDTree or BallTree) based on the dataset characteristics. Minkowski distance was used as the distance metric, generalising both Euclidean and Manhattan distances. The hyperparameters are shown in Table V.

TABLE V: Hyperparameters for K-Nearest Neighbors

| Hyperparameter | Value |
|---|---|
| Number of Neighbors (k) | 5 |
| Weights | Uniform |
| Algorithm | Auto |
| Distance Metric | Minkowski |

While KNN is computationally efficient during training (as it stores the training data directly), it can be slower at inference time, particularly with large feature sets or datasets. In our evaluation, inference times remained within acceptable limits due to the moderate size of the test dataset.

## D. Model Training and Evaluation Setup

All ML models were implemented using the `scikit-learn` framework in Python. The dataset was split into training (70%) and testing (30%) sets using stratified sampling to maintain class balance. All features were scaled using standardisation to ensure uniform treatment across algorithms. Model performance was assessed using standard classification metrics: accuracy, precision, recall, and F1-score. Hyperparameters were optimised using grid search with 5-fold cross-validation on the training set. The next section presents the experimental results along with an analysis of classifier performance and key observations across IoT device categories.

## V. Device Functionality Identification via SNORT Rules

Beyond device classification, network traffic patterns revealed identifiable functional behaviours across IoT categories, independent of manufacturer. For instance, smart home hubs commonly communicate with specific servers depending on active functions. Examples include querying news servers for updates or streaming music from proprietary services. These recognizable traffic signatures enable the creation of SNORT intrusion detection system (IDS) rules to automate the detection of device actions.

Table VI illustrates a basic SNORT rule structure for monitoring TCP traffic from or to specific IP addresses. The `alert` keyword triggers an alert upon a rule match, with any and `tcp` specifying the protocol scope of the rule.

TABLE VI: SNORT IP Alert Rule Structure

| SNORT Rule Component | Description |
| --- | --- |
| alert | Generates an alert when the rule conditions are met |
| any | Applies to packets from or to the specified IP address |
| tcp | Specifies monitoring of the TCP protocol |
| IP Address | Target IP address to monitor |

Alerts are generated when connections are established with IP addresses stored in a manually curated database, which can be expanded over time. Firewall rules based on these alerts monitor DNS queries, as shown in Table VII, which uses regular expressions (REGEX) or specific IP addresses to detect DNS traffic associated with known IoT servers.

TABLE VII: SNORT DNS Alert Rule Structure

| SNORT Rule Component | Description |
| --- | --- |
| alert | Generates an alert on matching conditions |
| any | Applies to any UDP packet on the specified port |
| udp | Specifies monitoring of the UDP protocol |
| Port | Target port for monitoring DNS queries |
| REGEX / IP Address | Pattern or address triggering the alert |

These SNORT rules enable monitoring of DNS request frequencies and destinations, which can reveal manufacturer-specific domains and facilitate device brand identification.

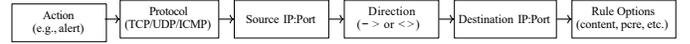

Fig. 2: Structure of a Custom Snort Rule

Such visibility enhances network security by enabling administrators to monitor IoT device activity patterns and detect potentially unauthorised or anomalous behaviours.

Figure 2 illustrates the sequential structure of a custom Snort rule. The rule begins with the alert keyword specifying the action to perform, followed by the any keyword indicating that the rule applies to any source or destination. The protocol field specifies whether the rule monitors TCP or UDP traffic. Next, the IP address and port number fields define the target network endpoint to observe. The pattern or regular expression (REGEX) field contains the specific signature or content to match in the packet payload. Finally, the rule specifies the generation of a log or alert, which enables network administrators to detect suspicious activities or unauthorised device behaviours in real-time.

## VI. Results and Analysis

The primary objective of this experiment was to investigate the feasibility of identifying IoT devices from outside a LAN by analysing WAN traffic and applying machine learning algorithms. Using a controlled network environment, traffic data was collected and analysed to classify device types and associated actions. Algorithms such as RF, KNN, and MLP were employed to identify patterns in the traffic, including IP addresses and DNS queries indicative of specific IoT devices. Building upon this analysis, an intrusion detection system (IDS) was developed using SNORT rules to monitor traffic for these identifiable patterns. This section presents the classification performance of each algorithm and demonstrates how specific device actions were detected using the developed IDS rules. These results offer important insights into the potential of WAN-based IoT monitoring for improving the security and observability of networked environments.

### A. Machine Learning Model Performance

TABLE VIII: Performance of Machine Learning Models

| Model | Accuracy (%) |
| --- | --- |
| Random Forest | 91.0 |
| Multilayer Perceptron | 56.0 |
| K-Nearest Neighbors | 79.2 |

As shown in Table VIII, the RF algorithm achieved the highest accuracy at 91%, significantly outperforming the other models. This indicates its strong capability in classifying IoT devices using WAN traffic data. The KNN model also performed reasonably well, achieving 79.2% accuracy. However, the Multilayer Perceptron, a type of artificial neural network, yielded significantly lower accuracy (56%), likely due to the relatively small dataset used. Deep learning models such as MLP generally require larger and more diverse datasets to perform optimally, and may not be suitable in data-constrained scenarios like this one.

## B. Device Action Detection

In addition to device classification, this experiment explored the detection of specific IoT device actions through traffic analysis. The success of this approach depended largely on the nature of the device and the encryption standards employed. For example, surveillance devices such as security cameras utilise modern TLS encryption, which significantly obscures their traffic and makes action detection more difficult. In contrast, other devices used outdated protocols such as SSL, making their communication patterns more accessible for analysis.

These differences in security implementation created both opportunities and limitations for action detection. Devices using weaker encryption exposed more identifiable traffic behaviour, while highly secure devices limited visibility into their activities. This highlights a fundamental trade-off between security and traffic observability in IoT systems.

TABLE IX: Detected Device Actions by Category

| Category | Devices | Detectable Actions |
|---|---|---|
| Hub | Amazon Echo, Google Home, Apple Home | Playing music or radio, accessing news content |
| Appliance | iRobot Vacuum, Sensibo AC Controller | Remote control via app or voice command |
| Streaming | Fetch Box Mini | Streaming cable television, user interface interaction |
| Energy | Smart Lighting Systems | Remote on/off control and brightness adjustment |

As shown in Table IX, functionality detection was possible for several devices, with actions categorised by device type. Notably, the detection of specific behaviours was more feasible for hubs and entertainment devices, whose traffic included regular connections to known content servers. In contrast, surveillance devices offered limited visibility due to their use of robust encryption.

A key limitation of the detection process was the reliance on prior traffic analysis to identify action patterns. For untested or newly introduced devices, action classification was not immediately feasible without additional data. This underscores the need for adaptive and scalable methods that can generalise to unseen devices without relying heavily on manual rule creation or extensive prior profiling.

## VII. CONCLUSION

This study presents a proof-of-concept framework for identifying IoT devices and their associated actions through the analysis of external network traffic. The findings demonstrate the feasibility of using machine learning (ML) techniques to classify device types and infer operational behaviours, thus addressing both active threats and passive surveillance attacks in IoT-enabled environments. This highlights the critical importance of proactive IoT monitoring as a component of broader network security strategies. Despite these promising results, several limitations warrant consideration. The dataset used for training was constrained in both size and diversity, as all traffic traces were manually generated within a controlled testbed. The absence of large-scale, publicly available IoT traffic datasets limited the generalisability of our models. Expanding the dataset through continuous traffic collection across a broader range of devices and operational contexts would improve classification accuracy and robustness. Furthermore, future work may explore advanced deep learning models capable of handling previously unseen devices and more complex traffic patterns.

Another avenue for enhancement involves automating the creation of intrusion detection rules, such as Snort alerts, for identifying device-specific actions. The current manual rule generation process restricts scalability and operational deployment. Integrating automated signature extraction and rule generation techniques would enable real-time monitoring and large-scale application of this proof-of-concept across diverse network environments. In summary, this research establishes a foundation for ML-based IoT device and action identification using network traffic analysis from an external perspective. With further refinement of datasets, the adoption of advanced ML models, and automated IDS integration, this approach can evolve into a practical and scalable solution for securing IoT-dense networks against emerging cyber threats.

REFERENCES

[1] W. Ahmad, M. Al-Rakhami, A. Gumaei, M. S. Hossain, and M. M. Su'ud, "Sociotechnical analysis of factors influencing IoT adoption in healthcare: A systematic review," *Technology in Society*, vol. 77, p. 102430, 2024.
[2] Kumar, Saurav, and Ajit kumar Keshri. "An effective DDoS attack mitigation strategy for IoT using an optimization-based adaptive security model." *Knowledge-Based Systems*, Vol 299, pp. 112052, 2024.
[3] S. Gvozdenovic, J. K. Becker, J. Mikulskis, and D. Starobinski, "IoT-Scan: Network Reconnaissance for Internet of Things," *IEEE Internet Things J.*, vol. 11, no. 8, pp. 13091–13107, Apr. 2024.
[4] H. Hindy *et al.*, "Machine Learning Based IoT Intrusion Detection System: An MQTT Case Study (MQTT-IoT-IDS2020 Dataset)," in *Lecture Notes in Networks and Systems*, vol. 180, B. Ghita and S. Shiaeles, Eds. Cham: Springer, 2021, pp. 79–92, doi: 10.1007/978-3-030-64758-2_6.
[5] D.-R. Berte, "Defining the IoT," in *Proc. Int. Conf. Business Excellence*, vol. 12, no. 1, pp. 118–128, May 2018, doi: 10.2478/picbe-2018-0013.
[6] P. Bajpai, A. Sood, and R. Enbody, "The Art of Mapping IoT Devices in Networks," *Network Security*, vol. 2018, no. 4, pp. 8–15, Apr. 2018.
[7] M. Niedermaier, F. Fischer, D. Merli, and G. Sigl, "Network Scanning and Mapping for IIoT Edge Node Device Security," in *Proc. Int. Conf. Applied Electronics (AE)*, pp. 1–6, Sept. 2019.
[8] L. Zhu, J. Gao, W. Yang, and Y. Xu, "Edge Intelligence for Internet of Things: A Survey," *IEEE Access*, vol. 9, pp. 54574–54617, 2021, doi: 10.1109/ACCESS.2021.3065123.
[9] A. De Resende, P. De Melo, J. Souza, R. Cattelan, and R. Miani, "Traffic Classification of Home Network Devices using Supervised Learning," in *Proc. 14th Int. Conf. Agents and Artificial Intelligence*, vol. 3, pp. 114–120, Feb. 2022, doi: 10.5220/0010785500003116.
[10] Y. Liu, J. Wang, J. Li, S. Niu, and H. Song, "Machine Learning for the Detection and Identification of Internet of Things (IoT) Devices: A Survey," 2021. [Online]. Available: arXiv:2101.10181.
[11] A. Sivanathan, "IoT Behavioral Monitoring via Network Traffic Analysis," Ph.D. dissertation, Sch. Elect. Eng. & Telecommun., Univ. New South Wales, Sydney, 2019. [Online]. Available: arXiv:2001.10632.
[12] L. Santos, C. Rabadao, and R. Goncalves, "Intrusion Detection Systems in Internet of Things: A Literature Review," in *Proc. 13th Iberian Conf. Inf. Syst. Technol. (CISTI)*, Jun. 2018.
[13] Google, "IPv6 Adoption," accessed Aug. 6, 2025. [Online]. Available: https://www.google.com/intl/en/ipv6/statistics.html.
[14] M. Rani, A. George, and N. Muraleedharan, "Network Traffic Classification Using Supervised Learning Algorithms," in *Proc. Int. Conf. Computer, Electrical & Communication Eng. (ICCCECE)*, 2023, doi: 10.1109/ICCECE51049.2023.10084931.